\documentclass[graybox]{svmult}

\usepackage{helvet}         
\usepackage{courier}        
\usepackage{type1cm}        
\usepackage{makeidx}         
\usepackage{graphicx}        
\usepackage{multicol}        
\usepackage[bottom]{footmisc}

\usepackage{cite,url}
\usepackage{amssymb,amsmath}

\usepackage{fourier-orns}

\newcommand\crulefill[1][1ex]{\leavevmode\leaders\hrule depth \dimexpr-#1+0.4pt height #1\hfill\kern0pt}

\def\IC{\mathbb{C}}

\def\IP{\mathbb{P}}

\def\cD{{\mathcal D}}

\def\cV{\mathcal{V}}

\def\cT{\mathcal{T}}
\def\cD{\mathcal{D}}

\newtheorem{observation}{\bf OBSERVATION}

\begin{document}

\title*{From the String Landscape to the Mathematical Landscape: \\
a Machine-Learning Outlook}
\titlerunning{Machine-Learning Mathematics}
\author{Yang-Hui He}
\institute{Yang-Hui He \at 
London Institute for Mathematical Sciences, 
Royal Institution of Great Britain, 21 Albemarle St., London, W1S 4BS;
Merton College, University of Oxford, OX1 4JD, UK;
Department of Mathematics, City, University of London, EC1V0HB, UK;
School of Physics, NanKai University, Tianjin, 300071, P.R. China.\\
\email{hey@maths.ox.ac.uk}}
%
%
\maketitle

\abstract{We review the recent programme of using machine-learning to explore the landscape of mathematical problems.
With this paradigm as a model for human intuition - complementary to and in contrast with the more formalistic approach of automated theorem proving - 
we highlight some experiments on how AI helps with conjecture formulation, pattern recognition and computation.
}

\section{The String Landscape}
Perhaps the greatest theoretical challenge to string theory as a theory of everything is the vast proliferation of possible vacuum solutions, each of which is a possible 4-dimensional ``universe'' that descends from the 10 spacetime dimensions of the superstring.
This is the so-called ``vacuum degeneracy problem'', or the ``string landscape problem".
The reason for this multitude is the vast number of possible geometries for the missing 6 dimensions.
Whether we consider compactification, where the a Calabi-Yau manifold constitutes the missing dimensions,
or configurations of branes whose world-volumes complement these dimensions, we are inevitably 
confronted with the heart of the problem: geometrical structures, often due to an underlying combinatorial problem, tend to grow exponentially with dimension.

We can see this from estimates of possible vacua, which engender such astronomical numbers as
$10^{500}$ to $10^{10^5}$ \cite{Kachru:2003aw,Taylor:2015xtz,Halverson:2017ffz}.
These estimates come from tallying ``typical'' number of topologies of ``typical'' manifolds, as governed by the number of holes (or more strictly, algebraic cycles) of various dimensions within the manifolds.
Such topological quantities are immanently combinatorial in nature.

Lacking a fundamental ``selection principle'' \cite{Candelas:2007ac} - which would find {\it our universe} among the myriad -  the traditional approaches have been statistical valuations \cite{Douglas:2003um}, or brute-force searching for the Standard Model \cite{Braun:2005nv,Bouchard:2005ag,Gmeiner:2005vz,Anderson:2011ns,Cvetic:2020fkd,Constantin:2018xkj} as a needle in the haystack.
Whilst these approaches have met some success, the overwhelming complexity (especially in the computational sense \cite{Halverson:2018cio}) of, and the want of a canonical measure \cite{He:2021eiu} on, the string landscape, beckon for a paradigmatically different method of attack.

As the Zeitgeist of Artificial Intelligence (AI) breathes over all disciplines of science \cite{DL} in recent times, and as we firmly enter the era of Big Data and Machine-Learning (ML), it is only natural that such a perspective be undertaken to explore the string landscape.
This was indeed done in 2017 when ML was introduced into string theory \cite{He:2017aed,He:2017set,Krefl:2017yox,Carifio:2017bov,Ruehle:2017mzq}.
In particular, the proposal of \cite{He:2017aed} was to see whether ML could be used to study the databases in algebraic geometry, which have been compiled over the last few decades for the sake of studying string theory in physics and concepts such as mirror symmetry in mathematics.
To some details of this programme let us now turn.

\subsection{Calabi-Yau Manifolds: from Geometry to Physics}

\begin{figure}[h]
\sidecaption[t]
\includegraphics[scale=0.23]{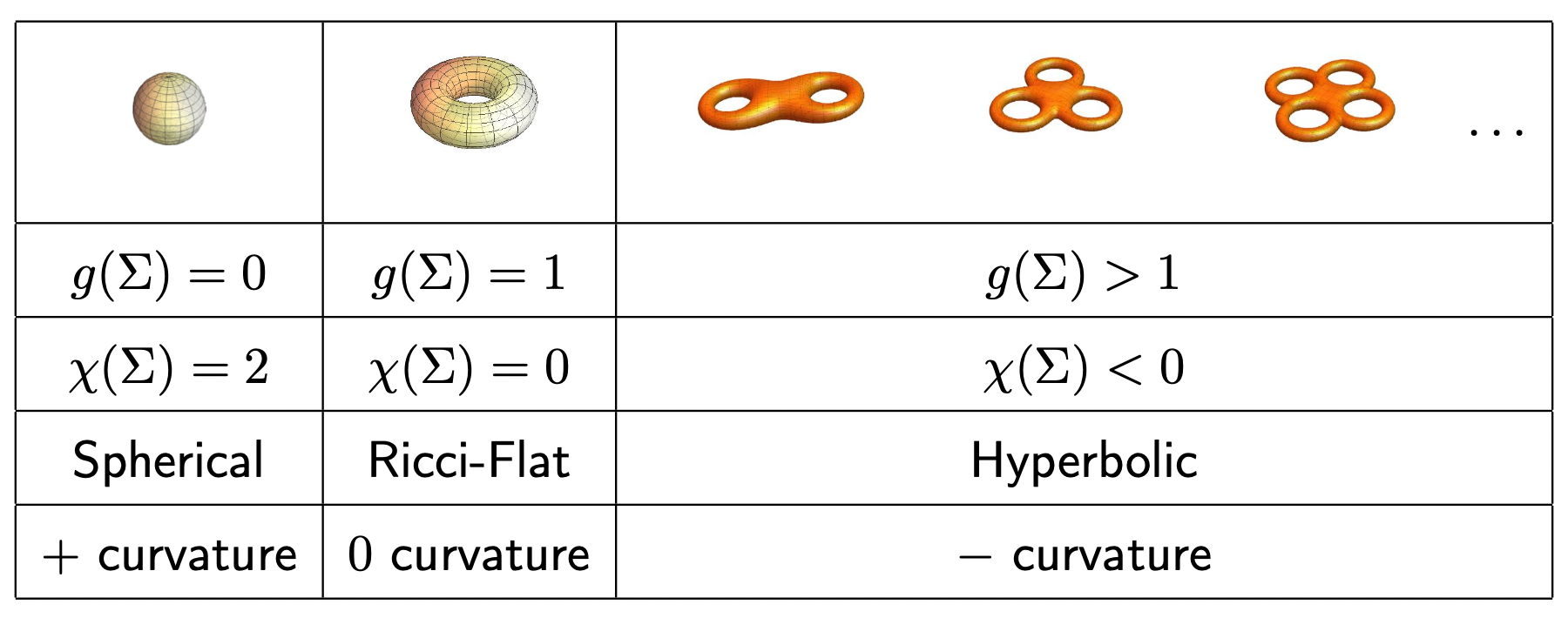}
\caption{The trichotomy of (smooth, compact, boundary-less) surfaces, organized according to toplogical type and related curvature.}
\label{f:surface}
\end{figure}

The classification of (compact, smooth, boudary-less) surfaces $\Sigma$ goes back to at least Euler, who realized that a single integer, called {\it genus}, completely characterizes the topological type of $\Sigma$.
Roughly, the genus $g$ counts the number of ``holes'': a sphere $S^2$ has genus 0, a torus $T^2 = S^1 \times S^1$ has genus 1, etc.
The {\it Theorema Egregium} of Gauss then relates topology to metric geometry: 
\begin{equation}\label{gauss}
2 - 2g = \frac{1}{2\pi} \int_\Sigma R \ .
\end{equation}
In the above, the combination $\chi = 2-2g$ is the Euler number and $R$ is the (Gaussian) curvature.
We therefore see a natural trichotomy of surfaces, as summarized in Fig.~\ref{f:surface}: negative, zero and positive curvature, with the boundary case of $R=0$, or Ricci-flatness, being the torus $T^2$.

\begin{figure}[h]
\sidecaption[t]
\includegraphics[scale=0.2]{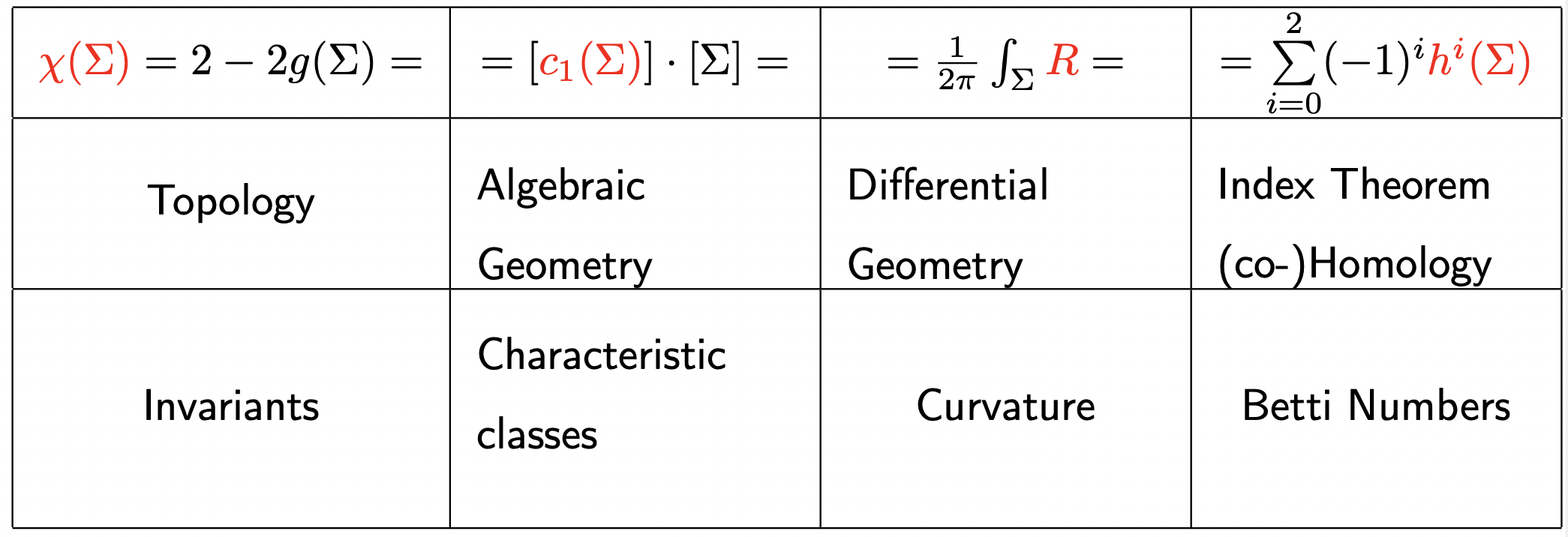}
\caption{The index theorem relating differential/algebraic geometry/toplogy for surfaces as complex algebraic curves.}
\label{f:index}
\end{figure}

With Riemann enters complex geometry: $\Sigma$ is not merely a real dimension 2 manifold, but a complex dimension 1 manifold. The trichotomy in this context manifests as Riemann Uniformization.
Complexification allows us to employ the powers of algebraic geometry over $\IC$ and $\Sigma$ can thus be realized as a complex algebraic curve. For instance, it can be the vanishing locus of a complex polynomial in the three projective variables $[x:y:z]$ of $\IC\IP^2$. The torus, in particular, can be realized as the famous cubic elliptic curve.
In modern parlance, the Gaussian integral is thought of as the intersection theory between homology (the class $[\Sigma]$) with cohomology (the first Chern class $c_1(\Sigma)$).
Likewise, $\chi$, by the index theorem, is the alternating sum of dimensions of appropriate (co-)homology groups. We summarize this beautiful story, spanning the two centuries from Euler to Chern, Atiyah, Singer et al., in Fig.~\ref{f:index}.

Generalizing Figs.~\ref{f:surface} and \ref{f:index} to complex dimension higher than 1 is, understandably, difficult. However, at least for a class of complex manifolds, called K\"ahler, whose (Hermitian) metric $g_{\bar{\mu}\nu}$ comes from a single scalar potential $K$ as $g_{\bar{\mu}\nu} = \partial_{\bar{\mu}} \partial_{\nu} K$, the story does extend nicely: the Chern class governs the curvature.
This is roughly the content and significance of the Calabi conjecture \cite{calabi}, which Yau proved some 20 years later in his Fields-Medal-winning work \cite{yau}.

It is serendipitous that when string theorists worked out the conditions for compactification in the incipience of string phenomenology \cite{Candelas:1985en}, one of the solutions (and today still standard) for the extra 6-dimensions is a complex, K\"ahler, Ricci-flat 3-fold. Furthermore, Strominger, one of the authors, was Yau's visitor at the IAS. And thus the world of high-energy theoretical physics intermingled with the world of complex algebraic geometry.
In fact, the physicists named such manifolds ``Calabi-Yau'' (CY), and the rest, was history.
The torus $T^2$, is thus a premium example of a Calabi-Yau 1-fold, of complex dimension 1.
The reader interested in further details of the Calabi-Yau landscape as a confluence between physics, mathematics and modern data science, is referred to the pedagogical book \cite{He:2018jtw}.

Over the decades since the mid-1980s, a host of activity ensued in creating large data-bases of CY manifolds for the intention of sifting through to find the Standard Model. Perhaps it was unexpected that the number \footnote{
By contrast, a CY 1-fold can only be $T^2$, a CY 2-fold can only be $T^4$ and K3. We therefore see the aforementioned exponential growth of possibilities as we increase in dimension.
Nevertheless, it is a standing conjecture of Yau that the number of possible topological types of CY in every dimension is {\it finite}.
} of CY 3-folds reached billions by the turn of the century (and still growing!) \cite{Kreuzer:1995cd}.
Furthermore, the sophisticated machinery of modern geometry, much of which was inherited from the Bourbaki School, was used to compute the various quantities (q.v.~the classic \cite{hartshorne} and for physicists, \cite{hubschbook}), particularly the topological ones such as Euler, Betti and Hodge numbers, which have precise interpretation as  Standard Model particles.

\subsection{Machine-Learning Algebraic Geometry}
The {\it point d'appui} of \cite{He:2017aed} was that these large sets of CY manifolds constituted labelled datasets ripe for machine-learning. In fact, the situation is even more general and {\it any} mathematical computation can be thought of this way. We shall not delve into the details of CY topological invariants or string phenomenology, but the idea can be construed as follows.
The purpose of algebraic geometry is to realize a manifold as the vanishing loci of a system of multi-variate polynomials where the variables are the coordinates of some appropriate ambient space such as projective space. We can thus represent a manifold as a list (tensor) of coefficients \footnote{
These coefficients determine the ``shape'' of the manifold. In {\sf Mathematica}, there is a convenient command for this, viz., {\sf CoefficientList[~]}.
}. Traditional methods such as exact sequences and Gr\"obner bases (q.v.~\cite{mike} for ML on selecting S-pairs) then computes desired geometrical quantities such as  Hilbert series or Betti numbers.
In the special case of extracting topological quantities, the coefficients are irrelevant (topology does not depend on shape) and we have even simpler representations. For instance, one could record just the degrees of the various defining polynomials.

But a tensor can naturally be interpreted as a pixelated image (up to some normalization and padding if necessary), and thus the general statement of \cite{He:2017aed,He:2017set} is that
\begin{observation}
Computation in algebraic geometry is an image-recognition problem.
\end{observation}
To make this observation concrete, let us give an example.
Suppose we are given a CY 3-fold \footnote{
Strictly, this is a family of manifolds since we are not specifying the coefficient which dictate complex structure (shape).
}, defined by the intersection of 8 polynomials in a product $(\IC\IP^1)^6 \times (\IC\IP^2)^2$ of projective spaces given by the configuration below \footnote{
This is an example of a complete intersection CY in product of project spaces (CICY), which was possibly the first database in algebraic geometry \cite{cicy}. To read it, each column is a defining polynomial. For example, the first column corresponds to a polynomial which is multi-linear in the first and second $\IC\IP^1$ factors and also linear in the first $\IC\IP^2$ factor.
}.
The topological quantity, a so-called Hodge number $h^{2,1}$ was computed (see \cite{hubschbook}) to be 22 using long exact sequence in cohomology induced by an Euler sequence (quite a difficult and expensive computation!). However, we could associate 0 to, say, purple, green to 1 and red to 2. After padding with 0 (to normalize over the full CY dataset of which this is one case), and the computation of $h^{2,1}$ becomes an image-processing problem no different than hand-writing recognition:
\begin{equation}\label{h21}
h^{2,1}({\tiny\left(\arraycolsep=1.4pt\def\arraystretch{0}
\begin{array}{cccccccc}
 1 & 1 & 0 & 0 & 0 & 0 & 0 & 0 \\
 1 & 0 & 1 & 0 & 0 & 0 & 0 & 0 \\
 0 & 0 & 0 & 1 & 0 & 1 & 0 & 0 \\
 0 & 0 & 0 & 0 & 1 & 0 & 1 & 0 \\
 0 & 0 & 0 & 0 & 0 & 0 & 2 & 0 \\
 0 & 1 & 1 & 0 & 0 & 0 & 0 & 1 \\
 1 & 0 & 0 & 0 & 0 & 1 & 1 & 0 \\
 0 & 0 & 0 & 1 & 1 & 0 & 0 & 1 \\
\end{array}
\right)}) = 22
\mbox{ \qquad becomes \qquad }
\begin{tabular}{c}\includegraphics[trim=0mm 0mm 0mm 0mm, clip, width=0.5in]{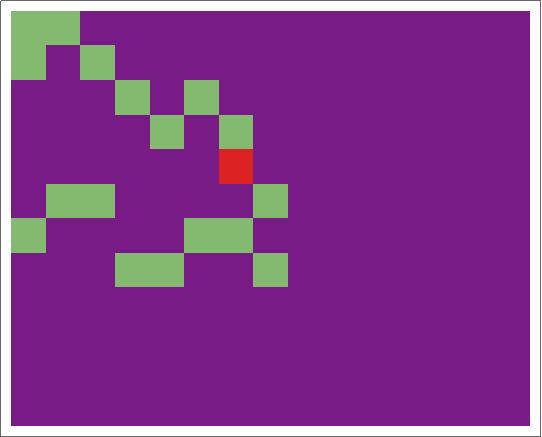}\end{tabular}
\longrightarrow
22 \ .
\end{equation}

A surprising result of \cite{He:2017aed} is that such labeled data, consisting of typically around $10^5 \sim 10^6$ points, when fed into a standard ML algorithm, such as a fairly shallow feed-forward neural network (otherwise known as an MLP) with sigmoid activation functions, or a support vector machine (SVP), achieves over 90\% accuracy in a standard 80-20 cross-validation \footnote{
In machine-learning, this means we take the full labelled data, train on 80\% randomly selected, and validate - meaning we check what the output is as predicted by the NN versus the actual value - on the {\it unseen} 20\%.
} in a matter of seconds on an ordinary laptop.
Since then, more sophisticated neural networks (NNs) have achieved over 99.9\% accuracy \cite{Bull:2018uow,Krippendorf:2020gny,Erbin:2020tks,Altman:2021pyc} (q.v.~recognition of elliptic fibration within the data using ML \cite{He:2019vsj}).
How could a relatively simple ML algorithm {\it guess} at a cohomology computation, without any knowledge of the underlying mathematics?
At some level, this is the Universal Approximation Theorem of NNs at work \cite{UAT}, which states that at sufficiently large depth/width, a NN can approximate almost any map, much like the way a Taylor series can approximate any analytic function.
Yet, the relative simplicity of the architecture of the NN is highly suggestive of a method which bypasses the sophistication and computational complexity of the standard algorithms of algebraic geometry.
To this point let us now turn.

\section{The Landscape of Mathematics}
The great {\it utility} of our paradigm to the string landscape, and indeed to problems in theoretical physics, is obvious.
Even when not reaching 100\% accuracy, a rapid and highly accurate NN estimate could reduce practical computations, say, of searching the exact Standard Model within string string, many orders of magnitude faster.
Utility aside, the unexpected success of machine learning of algebraic geometry beckons a deeper question: can one machine learn mathematics \cite{He:2021oav,He:2018jtw}?
By this we mean several levels: can ML/AI (1) extract patterns from mathematical data, supervised and unsupervised, patterns which have not been noted by the human eye? (2) help formulate new conjectures and find easier formulae (q.v.~recent collaboration on how AI can help with mathematical intuition \cite{deepmind})? (3) help with new pathways in a proof? (4) help understand the structure of mathematics across the disciplines?

It is expedient to digress momentarily on some speculations upon the nature of mathematics whilst we are planning to explore Her landscape.
The turn of the 20th century witnessed a tension between two Schools of thought: (i) the logicism-formalism of Hilbert and (ii) the intuitionism-constructivism of Poincar\'e.
The first, rested in the tradition of Leibniz, Frege, Peano, Russell-Whitehead, Wittgenstein, et al., and attempted to logically build all of mathematics, without contradiction, symbol by symbol.
The second, propelled by Brouwer, Heyting, Poincar\'e et al., sought for a more ``human'' and experiential element to mathematics.

The advent of computers in mathematics has dawned a new era. More importantly, they are becoming more than a mere aid to computation.
There is a growing number of major results - championed by e.g., the 4-colour theorem, or the classification of finite simple groups - which {\it could not} have been possible without computer work. The reason is simple: the rate of growth of mathematical knowledge and the requisite length of many a proof have perhaps already exceeded the capacity of the human mind.
The full details of the proof of Fermat is hundreds of pages of highly technical mathematics understandable by a small community, that of the classification of simple groups, thousands.
It is entirely conceivable that the proof of the Riemann Hypothesis will take longer than several human lifetimes to construct or digest, even if we take into account the cumulative nature of research.

Consequently, Buzzard, Davenport et al. \cite{buzzard,ICM2018} have been emphasising how essential the Automated Theorem Proving programme (ATP) is to the future of mathematics. Software such as {\sf Lean} is currently constructing all statements and proofs of mathematics, symbol by symbol, line by line. Their optimistic estimate is that within 10 years, all of undergraduate level mathematics will be built from scratch automatically. More strikingly, some at Google Deepmind suspect that as computers defeated humans at chess in the 1990s and Go in the 2010s, they will beat us at producing new mathematics by 2030.

The ATP programme can be thought of as being along the formalistic skein of Hilbert, and, to borrow terminology from physics, one could call this ``bottom-up mathematics'' \cite{He:2021oav}.
Our foregoing discussion of using ML which attempts to extract patterns from data or extrapolate methods from heuristics, on the other hand, is much more along the intuitionistic line of Poincar\'e.
Again, to borrow from physics, one could call this ``top-down mathematics'' \cite{He:2021oav}.
These two threads should indeed be pursued in parallel and here, we shall summarize some recent experiments in the latter.

\subsection{Methodology}
For concreteness, let us focus on calculations of the form of \eqref{h21}, which should be ubiquitous in mathematics. We shall let $\cD := \{ T_i \to p_i \}$ be a set of input tensors $T_i$ with output property $p_i$, typically obtained from some exhaustive and intensive computation.
We then split $\cD := \cT \sqcup \cV$ into training set $\cT$ and validation set $\cV$ where $\cT$ is a random sample of, say, 80\%.
Such data, representing ``experience and intuition'' of the practitioner, could then be passed to standard machine-learning algorithms such as neural classifiers/regressors, MLPs, SVMs, decisions trees, etc. Importantly, these algorithms have {\it no prior knowledge} of the mathematics \footnote{
Of course, building activation functions which {\it know} some of the underlying theory is effective and computationally helpful, as was done in, e.g., \cite{Deen:2020dlf,gaussmanin,Douglas:2020hpv,Jejjala:2020wcc,Anderson:2020hux,Larfors:2021pbb,Gao:2021xbs}, but the true surprises lie in blind tests. This was performed in the initial experiments of \cite{He:2017aed} and the ones we shall shortly report, could lead to conjectures unfathomed by human thought.
}.
Once validation reaches high precision (especially 100\%), one could start formulating conjectures.
On the other hand, if one could not reach any good results exhausting a multitude of algorithms, it would indicate an inherent difficulty in the problem whence the data came.

\subsection{Across Disciplines}
With this method of attack it is natural to scan through the available data of mathematics, as a reconnaissance onto the topography of Her territory.
We saw in the above that algebraic geometry over $\IC$ responds well to ML and speculate that the reason for this is that all computations inherent thereto reduce to finding (co-)kernels of matrices.
Over the past 5 years, there have been various excursions into a variety of disciplines and we shall highlight some representative cases, and refer the reader to the citations as well as the summary in \cite{He:2021oav}.
~\\

\noindent {\bf Algebra: }
In \cite{He:2019nzx}, the question was posed as to whether one could ``see'' a finite group being simple or not, by direct inspection of its Cayley multiplication table. Surprisingly, an SVM  could do so to more than 0.98 precision, instigating the curious conjecture that simple and non-simple groups could be separable when plotting their flattened Cayley tables.
For continuous groups, the tensor decomposition into irreps for simple Lie algebras of type ABCD$G_2$ is computationally exponential as one goes up in weight. Yet, numerical quantities such as the number of terms in the decomposition can be quickly machine-learnt by an MLP with only a few layers to 0.96 precision \cite{Chen:2020jjw}.
In \cite{ideals}, MLPs, decision trees and graph NNs could distinguish table/non-table ideals to 100\% accuracy, whereby suggesting the existence of a yet-unknown formula.

\noindent {\bf  Graphs \& Combinatorics:  }
Various properties of finite graphs, such as cyclicity, genus, existence of Euler or Hamilton cycles, etc., were explored by ``looking'' at the the adjacency matrix with MLPs and SVMs \cite{He:2020fdg}.
The algorithms determining some of these quantities are quite involved indeed. For instance, Hamilton cycle detection is that of the traveling salesman problem, which is NP-hard. Typically, for these problems, one could reach 80-90\% accuracies, which could be related to the fact that detecting matrix permutations - and hence graph isomorphism - is currently a challenge to ML.
However, when more structures are put in, such as quiver representations \cite{Bao:2020nbi}, or tropical geometry \cite{Bao:2021olg}, accuracies in the high 90s can once more be attained.
Explorations in lattice polytopes \cite{Bao:2021ofk,Berglund:2021ztg} and knot invariants \cite{Gukov:2020qaj,Craven:2021ckk} also yield good results.

\noindent {\bf  Analytic Number Theory:  }
As one might imagine, uncovering patterns in arithmetic functions, such as prime characteristic, or the likes of Mobius $\mu$ and Liouville $\lambda$, would be very hard. And it turns out to be so not only for the human eye, but also for any standard ML algorithm \cite{He:2017aed,He:2018jtw,He:2021oav}.
Likewise, one would imagine finding new patterns in the Riemann zeta function \cite{zeta1,zeta} to be a formidable challenge.

\noindent {\bf  Arithmetic Geometry: }
Yet, with a mixture of initial astonishment and a posteriori reassurance, problems in arithmetic geometry are very much amenable to ML.
Properties such as the arithmetic of L-functions \cite{He:2020kzg,He:2020qlg}, degree of Galois extensions for dessins d'enfants \cite{He:2020eva}, or even the quantities pertaining to the strong Birch-Swinnerton-Dyer conjecture  \cite{Alessandretti:2019jbs,He:2020tkg} (interestingly, the most difficult Tate-Shafaverich group is the least responsive) can all be learnt to high accuracies.
Indeed, as exemplified by countless historical cases, translating Diophantine problems to geometry, especially that of (hyper-)elliptic curves, renders them much more tractable. In this sense, our ML methodology and results on the data are consistent with this notion that arithmetic geometry is closer to geometry than to arithmetic.

With these experiments, we conclude with the remark and speculation that there is a ``hierarchy'' of mathematical problems, perhaps in tune with our expectations: 
\begin{observation} Across the disciplines of mathematics,
\[
\begin{split}
\left[\mbox{numerical analysis}\right] < 
\left[\mbox{algebraic geometry over $\IC$ $\sim$ arithmetic geometry} \right] <  \\
\left[\mbox{algebra/representation theory} \right] < 
\left[\mbox{combinatorics} \right] < 
\left[\mbox{analytic number theory} \right]
\end{split}
\]
\end{observation}
where $a<b$ means patterns from problem from $a$ are more easily extractable than those from $b$, or indeed that problems in $a$ are more easily solvable.

Above all, we encourage the readers to take their favourite problems and data and see how well ML performs on them.

\hrulefill\hspace{0.2cm} \floweroneleft\floweroneright \hspace{0.2cm} \hrulefill

\begin{acknowledgement}
{\it Ad Musam Alphabeticam meam tenebris luscis siderium.} \\
The author is grateful to STFC UK for grant ST/J00037X/2, as well as the kind hospitality - virtual and in person (!) during 2021 - of 
Cambridge (Winter School on ML), Cairo (BSM 22), Sofia (Lie Theory in Physics XIV), Toulouse (Geometry, Topology \& AI), Lisboa (BH, BPS, \& QI), Trento (ML for HEP), Singapore (M-theory \& Beyond), Johannesburg (String Data 22), Bangalore (KAWS 22), Hang-Zhou, Pisa, Minnesota (IMA), Tokyo (NaapingClass), Leicester, and Galway, on the talks at which this review is based.
\end{acknowledgement}

\let\oldbibliography\thebibliography
\renewcommand{\thebibliography}[1]{\oldbibliography{#1}
\setlength{\itemsep}{0pt}} 

\end{document}